\documentclass[aps,pra,preprint,eqsecnum]{revtex4}

\begin{document}

\title{Probabilities as Measures of Information}

\author{Francis G. Perey}
\email[]{pereyfg@ornl.gov}

\affiliation{ORNL Corporate Fellow (retired)}

\date{2003 October 7}

\begin{abstract}
We analyze the notion that physical theories are quantitative and
testable by observations in experiments. This leads us to propose a
new, Bayesian, interpretation of probabilities in physics that
unifies their current use in classical physical theories,
experimental physics and quantum mechanics. Probabilities are the 
result of quantifying the domain of possibilities that results when 
we interpret observations within the framework of a physical theory. 
They could also be said to be measures of information used to make 
predictions based upon a physical theory.
\end{abstract}

\maketitle

\section{Introduction}
Even though the concept of
probabilities has been used in physics for more than 200 years,
Laplace and Gauss having made important contributions, there is no
general agreement today concerning their interpretation, even when
limited to their use in physics. There is not even a consensus among
physicists as to whether they are `logical' (`subjective', Bayesian)
or `physical' (`objective', statistical) when they are used in physics.
Probabilities do not enter in all our physical theories. They are
absent from classical mechanics, classical electromagnetic theory and
the theories of relativity. However, they play an essential role in
our most successful physical theory: quantum mechanics. Because 
there is complete agreement among physicists regarding how 
probabilities in physical theories should be tested via observations 
in experiments, the issue of why and/or how they arise in some 
physical theories and not in others is not thought to be fundamental 
to physics itself.
\\
For some physicists, probabilities in physical 
theories are `physical' in nature because there are some physical 
phenomena that are `stochastic' or `random.' For others, probabilities 
are `physical' in a `statistical' sense and they arise because of 
`incomplete' descriptions, as could be considered the case for classical statistical 
mechanics. For others still, different physical mechanisms would be 
responsible for the probabilities of classical physics and quantum mechanics.
\\
On the other side, during the last 50 years the 
notion that probabilities in physics were of a Bayesian nature has been getting 
more and more support. That this was the case for probabilities in classical physics is 
likely in large part due to Jeffreys[1]and Jaynes[2]. 
The notion that in experimental physics the concepts of Bayesian and statistical 
probabilities should be treated not only on an equal footing, but even numerically 
combined, has now received the imprimatur of the BIPM[3]. With the 
development of the field of quantum computation, during the last 10 years, 
the notion that the probabilities of quantum mechanics could also be Bayesian 
measures of information has been receiving some support (Caves et al[4], 
Mermin[5], Fuchs[6], Peres[7].) In our opinion, an essential obstacle 
to the general acceptance by the physics community that probabilities in physics 
are Bayesian in nature is the fact that they are said to be: ``degrees of belief that 
an event will occur'' as in the BIPM report[3], ``degrees of truth of 
an assertion conditional upon the truth of some other assertion(s),'' as 
advocated by Jaynes[2] following Cox[8], or similar 
concepts that are thought by most physicsts to lie outside the realm of physics.
\\
In section II, we show that
starting from the notion that classical mechanics is quantitative and
testable by observations in experiments we can generate the algebra
of the probabilities used in classical physics. Since there are
no probabilities in classical mechanics, we will obtain probabilities
when we deal quantitatively with the ambiguities that must result when,
given some observations in an experiment, we use classical mechanics
to make predictions for future observations that could be 
considered as tests of classical mechanics. Consequently, the 
probabilities so generated will be Bayesian in nature and 
can be said to be measures of the information we have when making 
predictions for the values of the observable dynamical variables 
of classical mechanics. In section III, we show how, due to a very 
well understood difference between observing
classical and quantum systems, classical mechanics must be modified
for quantum systems in order to be quantitative and this 
leads to the probability amplitudes and
to the principle of superposition of quantum mechanics. However, the
probabilities in quantum mechanics will have precisely the same
meaning as they have in our analysis of observations in experiments
using classical mechanics. They will quantify the ambiguities concerning 
predictions that we can make for future observations on quantum systems 
after we have made on them unambiguous observations. Therefore 
they will also be measures of the information we have when making 
predictions for the values of the observable dynamical variables of 
quantum mechanics.
\\
In our approach,
probabilities in physics will be thoroughly defined entities in terms
of well understood concepts of classical mechanics and now well
established mathematical tools. What we propose is a simple
explanation for why probabilities play a fundamental role in
classical and quantum physics, explanation that we think is missing
today. This explanation is not based upon any new physical mechanism.
It is based upon an interpretation of a well established mathematical technique
for dealing with a fundamental problem that experimentalists must face in
every experiment. Consequently, it could be said that our argument is
philosophical, rather than physical. While this does not present any
problem concerning classical mechanics, since it is well understood.
In the case of quantum mechanics, our interpretation of how and why
its probabilities arise leaves us in the
same conundrum as a plethora of other proposed interpretations that
do not affect in any way the current use of quantum
mechanics. However, what sets our argument apart from other
interpretations of the nature and meaning of probabilities in
physics, in particular in quantum mechanics, is that it rests upon a
simple statement that we think is non controversial and should be
acceptable to every physicist: `Physical theories are
quantitative and testable by observations in experiments.'
\section{Classical Mechanics}
We understand so well
classical mechanics that there seems to be no need to discuss at
length how classical mechanics is a quantitative theory: in its
formulas the symbols stand for real numbers. These formulas
will be numerically satisfied when we substitute into them the
numbers that correspond to the values of the measurable dynamical variables in a
given situation. Most frequently these formulas are used to make a
quantitative prediction for an observation corresponding to one of
the dynamical variable that was not measured in an experiment, on the
basis of the numerical values of the other dynamical variables that
were measured in that experiment. Quoting from a recent paper[9]:
``The description of the system can be given, in classical mechanics,
by a phase-space point. This point is the `true' point ---
others are false --- so the outcome of a measurement can
be predicted with certainty.'' We think that the overwhelming majority
of physicists, if not all of them, would agree with this statement.
The only trouble is that although we know that this statement is
theoretically correct, for the last two hundred years we have not
been able to do in practice, and will never be able to do, what we
state we should theoretically be able to do!
\\
It is frequently stated that
in contradistinction with quantum mechanics, classical mechanics is
deterministic, i.e. no probabilities enter in its formulas. However,
it is a fact that probabilities have been used for a long time to
convey the result of every measurement of the dynamical variables
that enter classical mechanics: position, volume, momentum, length,
etc... For some time now, no refereed physics journal will accept for
publication an experimental paper that will not include a quantified
discussion, using probabilities, of the measurements reported. Until
rather recently these probabilities were said in textbooks to
describe the `statistical errors' in the measurements, but that there
were also other experimental `errors' that were said to be
`systematic' in nature and these were not `quantified' by means of
probabilities because there were no `statistics' associated with 
them. Over the last thirty years or so, this point of view
has gradually changed, one no longer speaks of experimental `errors,'
whether `statistical' or `systematic,' but instead of experimental
uncertainties. Most significantly, the BIPM[3] now recommends  that
all measurement uncertainties, including those that are not
`statistical' in nature, be quantified by means of probabilities. The
probabilities used to quantify the systematic uncertainties are said
to correspond ``to a degree of belief that an event will occur.''
Although the BIPM report reviews at length the concept of
`statistical' probabilities, only two references to the concept of
`Bayesian statistics' are given and how they should be evaluated in
physics experiments is not discussed. Finally, the BIPM recommends
that all the uncertainties, whether `statistical' or `systematic' in
nature, be combined into a single probability statement.
\\
`Statistical' uncertainties
do not arise in many measurements of classical mechanics dynamical
variables. We therefore postpone discussing them until we have dealt
with the `systematic' uncertainties that are impossible to avoid.
Another reason we postpone discussing `statistical' uncertainties in
this paper is that the solution we propose for dealing with the
`systematic' uncertainties will give us the tool for dealing with the
`statistical' uncertainties. There is at least one reason why
`systematic' uncertainties are unavoidable, they must arise because
in order to be quantitative we make measurements and in every 
measurement we will specify the outcome with only a finite number of digits. We
are not at all concerned with the notion that some measurements may
be accurate enough for some specific purpose and consequently can be
considered exact for that purpose. What we analyze now is a fundamental problem that
results from the fact that in practice we must report a measurement
with a finite number of digits. How this arises can be made concrete
via the example of measuring the length of an object using a set of
`go, no-go' gauges. With a set of `go, no-go' gauges, if one repeats
the measurement, one always (i.e. systematically) gets the same
result: the value we seek is smaller than given by the `go' gauge but
larger than given by the `no-go' gauge, whatever are the units used
to label the set of gauges used. Clearly, these two different values
are not the length of the object, since it must be a single real number.
These two values are an attribute of the set of `go, no-go' gauges we use to make
the observation that we interpret as providing direct information
regarding the length of the object. If we had used a finer set of
`go, no-go' gauges we would likely have obtained two different
values,  but we would still have obtained two numbers. In the 
situation we have just described, there are no distribution of 
values that could be dealt with using statistical concepts.
\\
Many theoretical papers analyzing 
physical measurements in quantum mechanics 
view the interaction of the observer with the classical measuring 
instrument as the reading of a `pointer on the dial of a gauge'. Due to
the finite thickness of the tip of the pointer, and the finite size
of the dial of a gauge, one can only state with certainty that it
does not point below a certain value nor above another value, a
result which is conceptually identical to using a set of `go, no-go'
gauges. In our digital age, we could replace the traditional `pointer
on the dial of a gauge' with a `digital counter' that must
necessarily have a finite number of digits. We therefore know that
the `real number' we seek is not less than the value shown on the counter
but not more than the value obtained by increasing by 1 the least
significant digit. From a mathematical point of view, as opposed to a
practical one, the problem is the same whether we deal with real numbers
that have a single digit or many digits: there is a continuum of real numbers
between them. This source of `systematic' uncertainty is present in
every physics experiment. In many experiments there are other sources
of `systematic' uncertainties, but they are all of the same nature in
the sense that they provide both a lower and an upper bound to the
value sought. Of course, all sources of uncertainties in measurements
must be combined to yield an overall uncertainty. In this paper we do
not address this issue. In this section we analyze how one can deal
quantitatively with the problem of `systematic' uncertainties and we
further limit ourselves to analyze this problem in the context of a
classical mechanics experiment.
\\
Is one testing classical
mechanics when one reports the `measurement' of the length, the
position, the momentum, etc.. of an object to which classical mechanics applies? In
a very real sense yes because what is reported in experimental papers
as having been measured in an experiment, is the result of a process
that is often called the data analysis, or data reduction, where:
from the observations made in the experiment, using the relations of
classical mechanics, one makes a deduction concerning the value of
the quantity reported as having been measured. Therefore, in this
data analysis one is making a prediction on the basis of observations
made in the experiment and the quantitative relations of classical
mechanics in which the reported measured quantity enters and to which
the observations apply. If subsequent observations contradicted the
conclusions of the data reduction reported as the measured quantity
one would suspect either that a mistake was made or, very unlikely,
that classical mechanics was falsified in the experiment. Why should
probabilities appear in this process of data analysis since there are
no probabilities in the relations used in that data analysis? It is
not because experimentalists are uncertain concerning what they have
observed in an experiment. Any observation that could be said to be
uncertain should be categorically discarded in drawing predictions
based on the relations of classical mechanics. Fundamentally, the
reason why there are uncertainties in the predictions we make on the
basis of observations in experiments is that these predictions are
ambiguous. According to the relations of classical mechanics that
relate what was observed and what we seek to predict: several
different predictions, i.e. a range of values, would be consistent
with the observations, but according to classical mechanics that
prediction should be a unique real number. As discussed above, this must
necessarily occur if only because we must use a finite number of
digits.
\\
Are the notions of ``degrees
of belief that an event will occur'' or of ``degree of truth of an 
assertion conditional upon the truth of some other assertion(s)'' fully
acceptable in physics in order to assign a probability measure to the
possible values? We do not think so because, in our opinion, the
concepts of `degrees of belief' and `degree of truth' do not belong
in physics. In particular if, as we will show, what is accomplished
by using them can be achieved using concepts that have long been
considered essential in physics. Since physical theories are
quantitative, when faced with the problem of `systematic'
uncertainties, what we can do as physicists is be quantitative about
it and specify the number, or density, of different possible
predictions for the dynamical variable reported as having been
measured, given the unambiguous observations in the experiment and
the relations of classical mechanics that were used in making these
predictions. What we propose is to use a rationale and a mathematical
technique that is fundamentally different from the one used today to
quantify the `uncertainties' in the predictions, i.e. measurements.
Most of the time this new technique, as it should, will yield
results, i.e. probabilities, that are close, if not identical, to
those obtained today under a different justification. Therefore, what
we propose is a new logical and mathematical foundation for the
probabilities we use in classical physics.
\\
The proposed new technique is
based upon a very simple argument: since according to classical
mechanics the `value' (i.e. the real number) of the dynamical
variable we seek to predict, the length, the position, the volume,
etc.., is postulated to be unique in the experiment in question, what
we can do is formally transform this value, that we do not know but
should be one of the possible values, from the value it has to all
the values it has not: the other possible values. Because these
entities in classical mechanics are real numbers the transformations
in question will be additions or multiplications of real numbers and
their inverse. Since, given what was observed, every possible 
value could be the value it has
in the experiment, the transformations in question must be applied using
every possible value as the true value. Consequently, each of these transformations
has an inverse and one of them must be an identity transformation.
The transformations in question must therefore form a group. We refer
to this group as the possibilities generating transformation group for what will be
reported as the `measured' dynamical variable in the experiment. What
this does is immediately solve our problem of associating a quantitative measure
with the set of possibilities. In the case of classical mechanics,
the manifolds of the possibilities generating transformation groups are Euclidean
spaces, because the phase-spaces of classical mechanics are Euclidean
spaces. This is enough to establish that the algebra of a unique
measure associated with the set of possibilities will have the
algebra of probabilities of classical physics. This unique measure is
often called the `volume element' in the group manifold, the `Haar'
measure or the `weight function' in group space.
\\
We must emphasize that the
probabilities so generated are fully objective in the sense that any
one applying the same formulas of classical mechanics, given
precisely what was observed in the experiment must come up with
exactly the same probability measure for the set of possibilities.
Such types of probabilities are often said to be `subjective'
or Bayesian. However, these probabilities are not `objective,' in the
sense in which this term is used to specify a concept of
probabilities that would be `independent of any observer.' We must
emphasize that the concept of probabilities based upon the
possibilities generating transformation groups we propose requires two human
interventions: 1 - some observations must be made by someone, could
be made by someone or a situation is postulated to be; 2 - a physical
theory must be chosen to analyze the observations in question to 
quantify the number, or density, of possible values. When classical 
mechanics is the theory used in this analysis, there are no probabilities in 
that theory. We could say that the manifold of the possibilities generating transformation 
group for the value of a dynamical variable in an experiment represents our
state of knowledge of that dynamical variable in the experiment, given that
classical mechanics is used to analyze what was observed in the
experiment. Another way to look at the probabilities we have obtained
is that it is a measure associated with the set of possible values
that would be obtained if we were to make another, but more accurate,
measurement of that dynamical variable. To use a more modern 
terminology, we can say that the probabilities so obtained are a measure 
of the information we use to make a prediction based upon a physical theory 
and some observations.
\\
The notion that one could use group theory 
to formally deal with ambiguities is as old as group theory itself since at 
the time of his death Evariste Galois[10 ] stated that his main preocupation for 
some time had been the application of the theory of ambiguity to 
transcendental analysis, according to the concluding remark in his 
celebrated letter addressed to Auguste Chevalier two 
days before his death. The idea that Galois' theory of ambiguity was 
group theory and could be used to deal with uncertainties resulting 
from observations is not new and was pointed out by George 
Birkhoff[11 ] more than 60 years ago.
\\
 E. T. Jaynes[3] in the last 40 years has 
been a very eloquent champion of the point of view that probabilities 
were not physical, in particular in classical physics, including the use of 
transformation group methods. However, we must point out an essential 
and fundamental conceptual difference between what we have stated 
above and what Jaynes based his arguments upon. Jaynes 
defined probabilities as the ``degree of truth of an assertion 
conditional upon the truth of some other assertion(s)''. Then, to 
justify his transformation group method, he introduced a ``desideratum 
of consistency''[12 ]: ``In two different problems where we have
the same state of knowledge we should assign the same subjective
probability.'' Our objective is much more limited than the one
stated by Jaynes, we stay strictly within the ambit of physics, doing
what physicists have been doing for ages. Probabilities are not
introduced at the outset, they are a consequence of testing the
quantitative nature of our physical theories by observations in
experiments. They are the result of enumerating all the possibilities
according to a physical theory used in analyzing the observations
made in an experiment. The transformation groups used are based upon
the relations postulated in a specific physical theory and what is
observed in the experiment. The fraction of all the possibilities
that is associated with a subset of possibilities is what we call the
probability associated with that subset of possible values. This is of
course conditional upon having correctly identified what was
observed in the physical theory used and that this theory would not
be falsified in that experiment.
\\
So far our argument has been
completely abstract, we will now illustrate it with a few concrete
examples. First, we will deal with two problems that have been 
associated with concepts of probabilities since antiquity. However, we will 
analyze them as if they were classical physics experiments 
in which some observations are made and 
classical mechanics is used to quantify the ambiguities of predictions 
based upon the observations. Next, we will consider
some whose solutions using probabilities are more recent. Finally, we
will consider a well known 75 years old problem that, to our
knowledge, has so far resisted any satisfactory solution. We have two
further purposes in analyzing these classical mechanics problems. The
first one is that although we will be using group theory, it is not
because the problems we deal with have any `real' symmetry associated
with them, at best they could be said to present an `apparent'
symmetry, given what was observed and the theory used to analyze them.
The second one is that these
examples make very clear that the probabilities we generate are in no
way `physical' and the analysis of what was observed is based upon
the use of a specific physical theory.
\subsection{The Coin}
This
trivial example involves the information that a coin is at rest lying
flat on one side, which side is facing up is not observed. This
information, i.e. the data, could be the result of observing the coin
on edge, or a `friend' is telling us he observed the coin but does
not tell us which way is facing up. We seek to reduce the data we
have for which side is facing up. From classical mechanics, the
possibilities generating transformation group for which side is facing up has two
elements: a rotation of the coin of 180 degrees about a horizontal
axis changes which side is facing up and a rotation of 360 degrees
about the same axis is the identity transformation. This
transformation group leaves invariant the data: the coin is lying
flat on one side. Since this group is finite the same measure applies
to each transformation and therefore equal probability must be
assigned to each possibility. Obviously there is no symmetry in this
situation. The coin need not be in any way symmetrical,
nor need we be told that is was randomly placed, whatever this could
mean in a classical mechanics context. We also did not have to
consider ``consistent betting behavior'', or any ``Dutch-book argument'', for
which side is facing up, as is sometimes done (Caves et al[4]) to justify the
notion of Bayesian probabilities in physics. The instant our `friend', who 
could be standing next to us, tells us
which side of the coin is actually facing up, our possibilities
generating transformation group, which was different from his, `collapses' to its
subgroup: the identity transformation, a rotation of 360 degrees.
\subsection{The Playing Die}
This
second example is a little more interesting and deals with a
conventional right handed playing die. Let us consider that the
information, i.e. the data, we have concerning the die is that it is
at rest on an horizontal surface with one of its vertical side facing North. We 
reduce this data for the possible orientations of the die. We are dealing again
with a well known problem: the possibilities generating transformation group for the
orientation of the die is the octahedral group (O) with 24 elements.
We consequently obtain a probability of 1/24 for each possible
orientation of the die. Again, there is no real 
symmetry involved, the playing die in question need not be 
symmetrical nor do we need to be told that it was randomly tossed. 
If we wanted to we could reduce the same data only
for which side is facing up, ignoring the value on the side facing
North. In this case the possibilities generating transformation group is the
dihedral group ($D_3$) which has 6 elements. Each of the
possibilities is assigned a probability of 1/6. In the conventional
fashion we could call these probabilities 'marginal' since they
ignore which side is facing North. Finally, we could reduce the data
conditional upon a particular value on the side facing say North. The
possibility generating transformation group for the side facing up is the cyclic
group of order 4 ($C_4$). Consequently, we have in this case
a `conditional' probability of 1/4 for each of the possibilities for
the side facing up. As an important aside, `Bayes theorem' results
from possibilities generating groups that are the direct product of
two groups.
\subsection{When and How Fast}
We
now show, by considering the instant of time and interval of time
parameters of classical mechanics, how results that are often called
the `Laplace prior' and the `Jeffreys prior' can be obtained. Let
us say that in an experiment an event is observed to have occurred
not before a time $t_1$ and not after a time $t_2$.
Since the event must have occurred at a specific time $t$, we have the
data: $t_1 < t < t_2$. In classical mechanics
an instant of time is a continuous translation invariant parameter,
the possibilities generating transformation group for the time $t$ is the one
parameter Lie group $T_1$ where $t$ is transformed into $t'$
according to $t' = t + \alpha$ where $\alpha$ is the continuous
parameter of the group. The volume measure in the group manifold of
$T_1$ is constant. Therefore, the probability that the event
occurred in an interval of time $dt$ at any time $t$ between $t_1$
and $t_2$ is $dt/(t_2 - t_1)$. This result
is often called a Laplace prior: the probability distribution is
uniform in an interval.
\\
Next we look at the scale
parameter of the time-space of classical mechanics. Let us say that
in this case the period $\tau$ of a cyclic phenomenon is observed to
be not less than the time interval $\tau_1$ and no greater
than the time interval $\tau_2$. The data to be reduced is:
$\tau_1 < \tau < \tau_2$. Now the
possibilities generating transformation group for the period $\tau$ is the scale
transformation $\tau ' = \tau\cdot\beta$ where $\beta$ is the continuous
parameter of the group. The volume measure in the parameter space of
this group is no longer constant but proportional to $1/\beta$.
Therefore, the probability that the period of the cyclic phenomenon
in question is in the interval $d\tau$ about $\tau$ in the range 
$\tau_1 < \tau < \tau_2$ is: $d\tau /(\tau ln(\tau_2 / \tau_1))$
and zero elsewhere. This result is often called a Jeffreys prior. In
our proposed interpretation of the nature and meaning of
probabilities in physics, the qualifier `prior' to probabilities
should never be used since two things are always required in order to
assign probabilities: a physical theory and some observations. 

\subsection{von Mises' Water and Wine Problem}
One of the most interesting application by
Jaynes[13] of his group theoretical method is his proposed solution 
to the `Bertrand Chord Paradox'[14]. Bertrand's chord paradox is not
a physics experiment but one of plane geometry. Since we know all the
`quantitative relations' between the elements of the geometrical
figure involved, the groups of transformations to generate the
possibilities are well defined, as Jaynes shows. Jaynes group
theoretical method to solve `ambiguous' problems has been widely
criticized, see van Frassen[15], because of his inability to solve
or explain why he could not solve von Mises' water and wine problem.
As we will show, we also cannot solve von Mises's problem as he
stated it, but we can explain why and we can solve it if we consider
it a physics measurement problem.
\\
Von Mises[16]' original statement
of his water and wine problem is: ``We are given a glass
containing a mixture of water and wine. All that is known about the
concentrations of the liquids is that the ratio of water to wine is
not less than 1 and not more than 2; this means that the mixture
contains at least as much water as wine and at most, twice as much
water as wine.'' von Mises shows that if one applies Laplace's
prior first to the ratio of water to wine and then to the ratio of
wine to water one gets a contradiction. von Mises water and wine
problem has long been used as a litmus test for theories of rational
decisions in the face of uncertainty which, to our knowledge, none of
them has passed.
\\
Von Mises did not specify
what attribute of water and wine was used to express the
concentrations. If we are going to interpret this problem as one of
data reduction in a physics experiment this must be done. It is only
under such conditions that we could apply our theory of
transformations because we need to invoke the relations of the
physical theory used in which the given data enters. We will assume,
for simplicity, that in the theory used to analyze von Mises' data
the volumes are the attributes in question and that in the mixture
they are additive. Let us denote by $M$ the volume of the mixture in
the glass, by $E$ the volume of water in the mixture and by $V$ the
volume of wine in the mixture. Therefore, we consider that the
physical theory used to analyze the data states that $M = E + V$. From
which we obtain: $1 = f_e + f_v$ where $f_e$
is the fraction of the mixture which is water and $f_v$
the fraction which is wine. The given data is that $1 < (E/V) < 2$ 
from which we obtain:
\begin{equation} \label{eq:fv}
1/2 < f_e < 2/3
\end{equation}
Given that we have: $1 = f_e + f_v$ the possibilities generating 
transformation group for $f_e$, the
fraction of the mixture which is water, is the one parameter Lie
group $T_1$ where $f_e$ is transformed into
$f_e'$ according to $f_e' = f_e + \alpha$ and $\alpha$ is the continuous parameter
of the group. The volume measure in the group manifold of $T_1$
is constant. Of course, because the theory we use states that: $1 = f_e + f_v$, to 
any transformation of $f_e$
must correspond a transformation of $f_v$. Therefore,
the probability that the fraction of the mixture which is water is in
the interval $df_e$ about any value of $f_e$
in the interval: $1/2$ to $2/3$ is given by: $6 \cdot df_e$. It
is equally probable that $f_e$ is below or above $7/12$.
\\
We must emphasize that we
have not `solved' von Mises water and wine problem as he stated it.
We showed how to solve it if it was a physics experiment in which a
specific physical theory ( $M = E + V$ ) was used to analyze the
observations. Because wine contains alcohol, we have used an
incorrect `physical' theory to `solve' von Misses problem, we did so
on purpose to point out that the same data analyzed with different
theories could yield different probabilities. We do not know, and in
fact as physicist we do not care, if von Mises water and wine problem
can be solved as he stated it. As a referee for Physical Review we
would have requested that von Mises specifies what was `measured'
before accepting his `experiment' for publication.
\\
We could go on and derive
some well known results and distributions of classical physics[17].
As is the case for the Maxwellian and the normal distributions, it is
a very simple matter to derive them using possibilities generating transformation 
groups. As a matter of fact, Maxwell[18] hinself derived the Maxwellian 
distribution using an invariance argument. Most of 
Jaynes[2] derivations can be readily done
using transformation group techniques, since from the context one can
tell which physical `theory' and which `observations' are being used.
At this stage it is important to investigate whether our proposed
interpretation of the nature and meaning of probabilities in physics,
which yields mathematical results that have been used for over a
century in classical physics, is a step toward `understanding'
quantum mechanics.

\section{Non Relativistic Quantum Mechanics}
The obstacles to
understanding quantum mechanics can be appreciated when one realizes
that it is 75 years old and even though
many interpretations of it have been proposed, a definitive consensus
has not yet been reached (Zeilinger[19].) This in spite of the fact that
numerous international conferences have been devoted to the study of
its foundation. Such conferences have been held at least yearly for
more than 30 years without any apparent breakthrough. The situation
concerning understanding quantum mechanics has barely changed in the
40 years since Feynman[20], who played a major role in one of its
formulations, made his celebrated statement : ``I think I can
safely say that nobody today understands quantum mechanics.'' Most, if
not all, proposed interpretations of quantum mechanics focus upon
understanding its probabilities. Could our proposed interpretation of
the nature and meaning of probabilities in physics, based only upon
the notion that physical theories are quantitative and testable by
observations in experiments, provide a key to `understanding' quantum
mechanics? Only time will tell if what we propose is Rabi[21]'s `basic point' : 
``The problem is that the theory is too strong, too compelling. I feel we 
are missing a basic point. The next generation, as soon as they will have 
found that point, will knock on their heads and say: How could they 
have missed that?''
\\
There are many different
formulations of quantum mechanics. We do not think that at this stage
it is essential to proceed via an analysis of an axiomatization of
quantum mechanics. Because we claim to understand classical mechanics
and how the fact that it is quantitative and testable by observations
in experiments leads to the probabilities of classical physics, we
will focus upon a formulation of quantum mechanics that has a very
close and direct relationship to classical mechanics. It is one of
its earliest formulations: Dirac's Hamiltonian formulation of non
relativistic quantum mechanics. Quoting Dirac[22] (page 3) ``At this
stage it is important to remember that science is concerned only with
observable things and that we can observe an object only by letting
it interact with some outside influence. An act of observation is
thus necessarily accompanied by some disturbance of the object
observed. We may define an object to be big when the disturbance
accompanying our observation of it may be neglected, and small when
the disturbance cannot be neglected.'' Dirac then points out that one
can give an absolute meaning to size, the emphasis is Dirac's,:
``...we have to assume that {\sl there is a limit to the finiteness of
our powers of observation and the smallness of the accompanying
disturbance --- a limit which is inherent in the nature of things and
can never be surpassed by improved technique or increased skill on
the part of the observer.} If the object under observation is such
that the unavoidable limiting disturbance is negligible, then the
object is big in the absolute sense and we may apply classical
mechanics to it. If, on the other hand, the limiting disturbance is
not negligible, then the object is small in the absolute sense and we
require a new theory for dealing with it.'' Systems for which the
`limiting disturbance' is not negligible we will call quantum systems. 
Those system for which the `limiting disturbance' is negligible we will 
call classical systems.
\\
Why can we not use classical
mechanics for quantum systems? If classical mechanics was not
modified and taken to apply to a quantum system, the state of the
quantum system would be given by a point in an Euclidean phase space.
As explained in section II, even though we could
never experimentally prove it, our measurements should tell us that
the state of the system could be a point within a domain of phase
space that becomes smaller and smaller as we improve the accuracy of
the measurements on a particular system. But because of the well
understood manner in which a quantum system is disturbed when we
measure canonically conjugate dynamical variables, the above argument
breaks down completely when applied to quantum systems. The more
accurately we measure one of the observable dynamical
variable, the less accurately we can measure the conjugate dynamical
variable. This means that the state of a quantum mechanical system
cannot be given by a point in an Euclidean phase space. The concept
of a phase space is based upon the fact that in the relations of
classical mechanics we can take the symbols for conjugate dynamical
variables to stand for real numbers.
\\
The testability, by
observations in experiments, of the quantitative nature of a physical
theory rests upon the fact that in order to be measurable the
dynamical variables must be real numbers. This was fully understood
by Dirac[22] (page 34) ``When we make an observation we measure some
dynamical variable. It is obvious physically that the result of such
a measurement must always be a real number, so we expect that any
dynamical variable that we can measure must be a real dynamical
variable.'' To solve this problem Dirac invented his transformation
theory of non relativistic quantum mechanics in which the Hamiltonian
formulation of classical mechanics is preserved but the symbols for
the measurable dynamical variables are no longer directly real numbers, as they
are for classical systems, but Hermitian operators in an Hilbert
space. The eigenvalues of these Hermitian operators are the real
numbers needed to make the theory quantitative and testable by 
observations in experiments.
\\
Having obtained real numbers
for the measurable dynamical variables of quantum systems, as eigenvalues of
Hermitian operators, these can be measured
in experiments. This is done by having the quantum systems interact
with a classical system acting as the measuring instrument. We will
have to face the same problem that we had with measuring dynamical
variables of classical mechanics. That is to say, we will have
systematic uncertainties associated with their experimental
determination and these can be dealt with using possibilities
generating transformation groups to quantify their possible values. 
An interesting aspect of the fact that some Hamiltonians will lead to a 
discrete set of eigenvalues for a dynamical variable is that one then
does not need as high an accuracy in the classical measuring 
instrument to determine which state the quantum system is 
in, as is the case when we have a continum of eigenvalues. 
A consequence of having obtained the real numbers 
for the measurable dynamical variables of quantum systems via Hermitian 
operators is that we have eigenfunctions corresponding to
the eigenvalues. These eigenfunctions are rays in an Hilbert space
and strictly speaking there is nothing in classical mechanics that
corresponds to them. We are going to show, with the simplest possible
quantum system example, that these eigenfunctions can be interpreted as entities
that are completely analogous to the manifolds of the possibilities
generating transformation groups we introduced to deal quantitatively with
ambiguities that occur when we test classical mechanics in experiments.
\\
Let us consider having
observed, via a Stern-Gerlach experiment, an electron traveling along
the $y$-axis having its spin up in the $z$-direction. Our observation of
its spin up in the $z$-direction gives us, in an obvious basis, the
eigenfunction: 

\begin{equation}
|\Psi\rangle = { 1 \choose 0 }
\end{equation}

There is absolutely nothing
ambiguous in that eigenfunction concerning what would happen if we
were to repeat that measurement: the spin would be found to be up in
the $z$-direction. However, what does it tell us about the outcome of a
subsequent measurement of the spin in the $xz$ plane at an angle $\theta$
with the positive $z$-axis?
\\
The operator $\widehat S_\theta$
corresponding to that measurement can be expressed in terms of the
projection of the $x$ and $z$ operators:

\begin{equation}
\widehat S_\theta = sin(\theta) \widehat S_x + cos(\theta) \widehat S_z = 
\frac{\hbar}{2} {cos(\theta) \qquad sin(\theta) \choose sin(\theta) \quad -cos(\theta)}
\end{equation}

The eigenvalues of $\widehat S_\theta$ 
are $\hbar /2$ and $-\hbar /2$
and the corresponding eigenvectors are:

\begin{equation}
{cos(\theta/2) \choose sin(\theta/2)} \textrm{  and  } {-sin(\theta/2) \choose cos(\theta/2)}
\end{equation}

We can therefore write the
eigenfunction corresponding to our observation of the spin up in the
$z$-direction as:

\begin{equation}
{ 1 \choose 0 } = cos(\theta/2){cos(\theta/2) \choose sin(\theta/2)} - sin(\theta/2){-sin(\theta/2) \choose cos(\theta/2)}
\end{equation}

What this results precisely
means has been known for over 70 years. There are two possibilities:
the spin could be found to be up or down and we have an `amplitude'
associated with each possibility. This is the simplest example of the 
`superposition principle'. Because, the eigenfunction is a ray
in an Hilbert space, the `amplitude' associated with each
possibility is a complex number. In order to obtain a `real'
measure associated with each possibility we must use the `norm'
of these amplitudes. These norms are what
we call the probabilities for the various possibilities.
\\
Let us suppose we do perform the 
measurement of the spin observable at that angle $\theta$ The instant we 
observe which of the two possible eigenvalues prevails in this measurement, our 
state of knowlege concerning the value of the spin observable along 
that angle $\theta$ changes from ambiguous to certain. The eigenfunction 
which expressed the ambiguity concerning which eigenvalue would be 
observed collapses to the eigenfunction corresponding to the eigenvalue 
that was observed. This new eigenfunction can be used to make 
quantitative predictions for the spin values that could be observed in 
subsequent experiments.
\\
When we deal with classical 
systems the ambiguities concerning what could be observed in future 
experiments, given what was observed on this classical system, are 
expressed quantitatively via the Euclidean manifold of a possibilities 
generating transformation group. When we deal with a quantum system 
the same problem arrises but now the quantification of the ambiguities 
concerning what could be observed in the future is given by the ray in 
Hilbert space which is the eigenfunction corresponding to the eigenvalue 
of the dynamical variable that was observed. Such observations upon 
which we base predictions for future observations are often said today 
to prepare a quantum system in a given state. Consequently the `algebra' 
of the probabilities for quantum systems is different from the `algebra' of 
the probabilities for classical systems. This provides a complete explanation 
of Bell's inequalities which are applicable only to classical systems, 
and are indeed not applicable to quantum systems. From a logical point of 
view an even more fundamental difference between classical and quantum 
systems is their respective `principles of superposition'. In classical systems 
this principle refers to physical states of the systems. In quantum mechanics 
it refers to a state of ambiguity we have concerning predictions for future 
observations, when we have made some observations on , i.e. prepared, a 
quantum system.

\section{Conclusion}
We have shown that based upon
the requirement that physical theories be quantitative and testable
via observations in experiments, we can interpret probabilities in
both classical and quantum physics as being `logical' rather than
`physical'. Their Bayesian nature does not depend upon the
introduction of concepts such as `degrees of belief' or `degrees of
truth of assertions,' often associated today with the concept of
Bayesian probabilities, even in physics. They are the result of quantifying the domain
of possibilities that results when we interpret observations within
the framework of a physical theory. Probabilities in physics could
also be said to be measures of information interpreted within the
framework of a physical theory.
\\

\end{document}